\documentclass[prb,showkeys,twocolumn]{revtex4}

\usepackage{amsmath,amsfonts,amssymb,bm}
\usepackage{dcolumn}
\usepackage[final]{graphicx}
\usepackage{comment}
\usepackage{bm}

\begin{document}

\bibliographystyle{apsrev}

\title{Full three-dimensonal reconstruction of the dyadic Green tensor from electron energy loss spectroscopy of plasmonic nanoparticles}

\author{Anton H\"orl}
\author{Andreas Tr\"ugler}
\author{Ulrich Hohenester}\email{ulrich.hohenester@uni-graz.at}

\affiliation{Institute of Physics, University of Graz, Universit\"atsplatz 5, 8010 Graz, Austria}

\date{\today}

\begin{abstract}
Electron energy loss spectroscopy (EELS) has emerged as a powerful tool for the investigation of plasmonic nanoparticles, but the interpretation of EELS results in in terms of optical quantities, such as the photonic local density of states, remains challenging.  Recent work has demonstrated that under restrictive assumptions, including the applicability of the quasistatic approximation and a plasmonic response governed by a single mode, one can rephrase EELS as a tomography scheme for the reconstruction of plasmonic eigenmodes.  In this paper we lift these restrictions by formulating EELS as an inverse problem, and show that the \textit{complete} dyadic Green tensor can be reconstructed for plasmonic particles of arbitrary shape.  The key steps underlying our approach are a generic singular value decomposition of the dyadic Green tensor and a compressed sensing optimization for the determination of the expansion coefficients.  We demonstrate the applicability of our scheme for prototypical nanorod, bowtie, and cube geometries.
\end{abstract}

\maketitle


\section*{Introduction}

Electron energy loss spectroscopy (EELS) is a powerful tool for the investigation of plasmonic nanoparticles.~\cite{garcia:10,kociak:14}  EELS is a technique based on electron microscopy and measures the probability of a swift electron to lose part of its kinetic energy through plasmon excitation as a function of electron beam position.  Following first proof of principle experiments,~\cite{bosman:07,nelayah:07} in the last couple of years EELS has been exhaustively used for the investigation of plasmon modes in single and coupled nanoparticles.


Despite its success, the interpretation of EELS data in terms of optical quantities, such as the photonic local density of states~\cite{novotny:06} (LDOS), remains challenging.~\cite{garcia:08,hohenester.prl:09}  To overcome this problem, in Ref.~\citenum{hoerl.prl:13} we formulated EELS as a tomography scheme~\cite{herman:80} and showed that under certain assumptions a collection of EELS maps can be used to reconstruct the three-dimensional mode profile of plasmonic nanoparticles.  A similar approach was presented independently by Nicoletti and coworkers,~\cite{nicoletti:13} who demonstrated the applicability of the scheme for a silver nanocube.  Extracting three-dimensional information through sample tilting was also shown for a split-ring resonator~\cite{voncube:14} and a nanocrescent using cathodoluminescence imaging.~\cite{atre:15} 

The problem with EELS tomography is that the measurement signal (the loss probability) is not simply the integral of local losses along the electron trajectory, but involves a two-step process where the swift electron first excites a particle plasmon and then performs work against the induced particle plasmon field.  This leads to a non-local response function, which allows for a tomographic reconstruction only under restrictive assumptions, such as the applicability of the quasistatic approximation or a plasmonic response governed by a single mode.  In this paper we use additional pre-knowledge, namely that the particle plasmon fields are solutions of Maxwell's equations and that the dyadic Green tensor~\cite{novotny:06} can be decomposed into modes, in order to rephrase EELS in terms of an inverse problem.  We develop a rather generic model for the EELS probabilities, which depends on a few parameters, and determine the parameters such that the model data match as closely as possible the measured data.  Within this approach we are able to obtain most accurate reconstructions of the dyadic Green tensor, which, in turn, allows us to extract the three-dimensional photonic LDOS from a collection of tilted EELS maps.  We demonstrate the applicability of our scheme for prototypical nanorod, bowtie, and cube geometries.


\section*{Theory}

We start by analysing EELS within a semi-classical framework,~\cite{garcia:10} where a swift electron propagating with velocity $\bm v$ loses a tiny part of its kinetic energy by performing work against the electric field $\bm E[\bm r_e(t)]$ produced by itself.  For sufficiently large velocities we can ignore velocity changes in the electron trajectory $\bm r_e(t)\approx\bm R_0+\bm vt$, with $\bm R_0$ being the impact parameter.  It is convenient to split $\bm E=\bm E_{\rm bulk}+\bm E_{\rm surf}$ into a bulk contribution~\cite{jackson:99} $\bm E_{\rm bulk}$, corresponding to the electric field within an unbounded homogeneous medium, and a surface contribution $\bm E_{\rm surf}$, corresponding to field modifications (including surface plasmons) from the interfaces between different materials.  Bulk losses are due to Cherenkov radiation and electronic excitations,~\cite{garcia:10} and the loss probability is obtained by simply multiplying the loss probability per unit length $\gamma_{\rm bulk}^j(\omega)$, inside material $j$ and for loss energy $\hbar\omega$, with the path length $\ell_j$ of the electron inside material $j$,
\begin{equation}\label{eq:bulk}
  \Gamma_{\rm bulk}(\omega)=\sum_j\gamma_{\rm bulk}^j(\omega) \ell_j\,.
\end{equation}
Bulk losses can be interpreted in terms of local scatterings where the electron emits a photon or excites electrons in the dielectric material, and loses part of its kinetic energies.  To compute the surface loss probability, we integrate the work $dW=e\bm E_{\rm surf}\cdot\bm v dt$ performed by the electron over the entire trajectory, and decompose it into the different loss energies $\hbar\omega$ according to
\begin{equation}
  W=e\int_{-\infty}^\infty {\bm v}\cdot \bm E_{\rm surf}[\bm r_e(t)]\,dt=
  \int_0^\infty \hbar\omega\,\Gamma_{\rm surf}(\omega)\,d\omega\,.
\end{equation}
Thus, the energy loss probability becomes~\cite{garcia:10}
\begin{equation}\label{eq:surf}
  \Gamma_{\rm surf}(\bm R_{\hat{\bm v}},\omega)=\frac e{\pi\hbar\omega}\int_{-\infty}^\infty \mbox{Re}\Bigl\{
  e^{-i\omega t}\bm v\cdot\bm E_{\rm surf}[\bm r_e(t),\omega]\Bigr\}\,dt\,,
\end{equation}
where we have explicitly indicated the dependence on the electron propagation direction and the impact parameter through $\bm R_{\hat{\bm v}}=(\hat{\bm v},\bm R_0)$.  To understand the physical process underlying Eq.~\eqref{eq:surf} it is convenient to introduce the current distribution $\bm J(\bm r,t)=-e\bm v\delta(\bm r-\bm r_e(t))$ of the swift electron and the dyadic Green tensor~\cite{novotny:06} $\bm G(\bm r,\bm r',\omega)$ that relates for a given frequency $\omega$ a current source at position $\bm r'$ to an electric field at position $\bm r$ via $\bm E(\bm r,\omega)=i\omega\mu_0\,\bm G(\bm r,\bm r',\omega)\cdot\bm J(\bm r',\omega)$.  The loss probability of Eq.~\eqref{eq:surf} can then be rewritten in the form
\begin{equation}\label{eq:surf2}
  \Gamma_{\rm surf}(\bm R_{\hat{\bm v}},\omega)=\frac{\mu_0}{\pi\hbar}\int\mbox{Im}\Bigl\{
  \bm J^*(\bm r,\omega)\cdot\bm G(\bm r,\bm r',\omega)\cdot\bm J(\bm r',\omega)\Bigr\}\,d\bm r d\bm r'
\end{equation}
where $d\bm r$ denotes integration over the spatial variable $\bm r$.  Contrary to Eq.~\eqref{eq:bulk}, the above expression describes a genuinely non-local self-interaction process where the electron first induces a field (through excitation of a surface plasmon) and then performs work against the induced field.

In Ref.~\citenum{garcia:08} the authors tried to interpret Eq.~\eqref{eq:surf2} in terms of the photonic local density of states~\cite{novotny:06} (LDOS)
\begin{equation}\label{eq:ldos}
  \rho_{\hat{\bm n}}(\bm r,\omega)=\frac{6\omega}{\pi\omega^2}
  \mbox{Im}\Bigl\{\hat{\bm n}^*\cdot \bm G(\bm r,\bm r,\omega)\cdot\hat{\bm n}\Bigr\}\,,
\end{equation} 
which is of paramount importance in the field of nanooptics and describes how the decay rate of a quantum emitter located at postion $\bm r$ and with dipole moment oriented along $\hat{\bm n}$ becomes modified in presence of a structured dielectric environment.  While such interpretation can be formally established for nano structures with translational symmetry along one spatial dimension, it becomes problematic for nanoparticles with generic shape.~\cite{hohenester.prl:09}

A different interpretation of Eq.~\eqref{eq:surf2} in terms of a tomography scheme was formulated independently in Refs.~\citenum{hoerl.prl:13,nicoletti:13}.  As a preliminary step, let us consider the bulk losses of Eq.~\eqref{eq:bulk} for a given $\bm R_{\hat{\bm v}}$ value.  Then, each point $\bm r$ inside a medium $j$ contributes with $\gamma_{\rm bulk}^j$ to the total loss rate.  Within the field of tomography~\cite{herman:80} it is well known that the three-dimensional profile of $\gamma_{\rm bulk}(\bm r)$ can be uniquely reconstructed from a \textit{sinogram} where bulk losses are recorded for all possible propagation directions $\hat{\bm v}$, using the inverse Radon transform.  Such tomography reconstruction is significantly more complicated for the surface losses of Eq.~\eqref{eq:surf2} since $\Gamma_{\rm surf}$ is not the sum of local losses (as in the bulk case) but governed by the self-interaction process of excitation and back-action.  Only for certain, rather restrictive simplifications a viable tomography scheme can be formulated:~\cite{hoerl.prl:13,nicoletti:13} the nanoparticles must be small enough such that the quasistatic approximation can be employed; the plasmonic response must be governed by a single plasmonic eigenmode; the sinogram must only consist of electron trajectories that do not penetrate the particle; the sign of the eigenmode potentials must be unique.  Although it has been demonstrated that reconstruction is possible in certain cases,~\cite{hoerl.prl:13,nicoletti:13} it is obvious that the above restrictions provide a serious bottleneck for general plasmon field tomography.

In this paper we formulate a significantly more general scheme, which approaches the reconstruction as an \textit{inverse problem} rather than a tomography scheme.  We first describe our approach, and discuss possible problems and generalizations at the end.  First, we decompose the dyadic Green tensor into a number of modes $\bm E_k(\bm r,\omega)$
\begin{equation}\label{eq:expansion}
  \bm G(\bm r,\bm r',\omega)\approx\sum_{k=1}^n C_k\, \bm E_k(\bm r,\omega)\otimes\bm E_k(\bm r',\omega)\,,
\end{equation}
where $C_k$ controls how much the different modes contribute to the decomposition.  In the following we only consider positions $\bm r$, $\bm r'$ outside the plasmonic nanoparticle and assume that $\bm E_k(\bm r,\omega)$ is a solution of Maxwell's equations.  The expansion of Eq.~\eqref{eq:expansion} is generally possible because $\bm G$ is a symmetric matrix that can be submitted to a singular value decomposition, with $C_k$ being the singular values and $\bm E_k$ the orthogonal matrices.  In this respect, Eq.~\eqref{eq:expansion} is similar to a wavefunction expansion in quantum mechanics into a complete set of basis functions. 

To be useful as a reconstruction scheme the modes $\bm E_k(\bm r,\omega)$ should be sufficiently well adapted to the problem such that a limited number $n$ suffices for a suitable representation of $\bm G(\bm r,\bm r',\omega)$.  Possible modes are quasi normal modes of the plasmonic nanoparticles,~\cite{sauvan:13,ge:14,maekitalo:14,alpeggiani:15} which have recently received considerable interest, or natural oscillation modes of our boundary element method approach (see Methods).  With these modes, the surface losses of Eq.~\eqref{eq:surf2} become
\begin{equation}\label{eq:surfn}
  \tilde\Gamma_{\rm surf}(\bm R_{\hat{\bm v}},\omega)\approx\frac{\mu_0e^2}{\pi\hbar}
  \sum_{k=1}^n \mbox{Im}\Bigl\{C_k\, A^+(\bm R_{\hat{\bm v}},\omega)A^-(\bm R_{\hat{\bm v}},\omega)\Bigr\}\,,
\end{equation}
where $A_k^\pm(\bm R_{\hat{\bm v}},\omega)=\int_{-\infty}^\infty e^{\pm i\omega z/v}\hat{\bm v}\cdot \bm E_k(\bm R_0+\hat{\bm v}z,\omega)\,dz$ is the averaged mode profile along the electron propagation direction.
We can now formulate our inverse problem as follows.  Suppose that one has measured EELS spectra $\Gamma_{\rm exp}$ for a given loss energy and for various impact parameters and electron propagation directions.  We then determine the coefficients $C_k$ such that the entity of measurement data differs as little as possible from the model data of Eq.~\eqref{eq:surfn},
\begin{equation}\label{eq:optim}
  \min_{C_k}\frac 12\left\|\Gamma_{\rm exp}(\bm R_{\hat{\bm v}},\omega)-
  \tilde\Gamma_{\rm surf}(\bm R_{\hat{\bm v}},\omega)\right\|_{L_2}^2\,,
\end{equation}
resulting in a least square optimization (we adopt the norm definitions $\|\bm x\|_{L_2}^2=\sum_i|x_i|^2$ and $\|\bm x\|_{L_1}=\sum_i|x_i|$).  Alternatively, in this work we will use a compressed sensing optimization~\cite{candes:08,yang:11} 
\begin{equation}\label{eq:compressed}
  \min_{C_k}\left[\Bigl\|C_k\Bigr\|_{L_1}+\frac 1{2\mu}\left\|\Gamma_{\rm exp}(\bm R_{\hat{\bm v}},\omega)-
  \tilde\Gamma_{\rm surf}(\bm R_{\hat{\bm v}},\omega)\right\|_{L_2}^2\right]\,,
\end{equation}
which attempts to minimize the moduli of the expansion coefficients, therefore the scheme is often referred to as a $L_1$-optimization, and $\mu$ is a parameter that allows to switch between genuine compressed sensing and least square optimizations.~\cite{yang:11}  For a sufficiently small number of expansion modes $\bm E_k$, the determination of the expansion coefficients $C_k$ is a highly overdetermined problem since the measured loss data can be assembled for many propagation directions and impact parameters $\bm R_{\hat{\bm v}}$ .  The only pre-knowlege entering our optimization is the self-interaction-type scattering process of the electron loss, Eq.~\eqref{eq:surf2}, and the assumption that the dynamics of the electric fields outside the plasmonic nanoparticles is governed by Maxwell's equations.  Importantly, once the coefficients $C_k$ are determined we have (approximately) reconstructed the dyadic Green tensor of Eq.~\eqref{eq:expansion}, which allows us to compute all electrodynamic properties including the photonic LDOS.


\section*{Results}

\begin{figure}
\centerline{\includegraphics[width=\columnwidth]{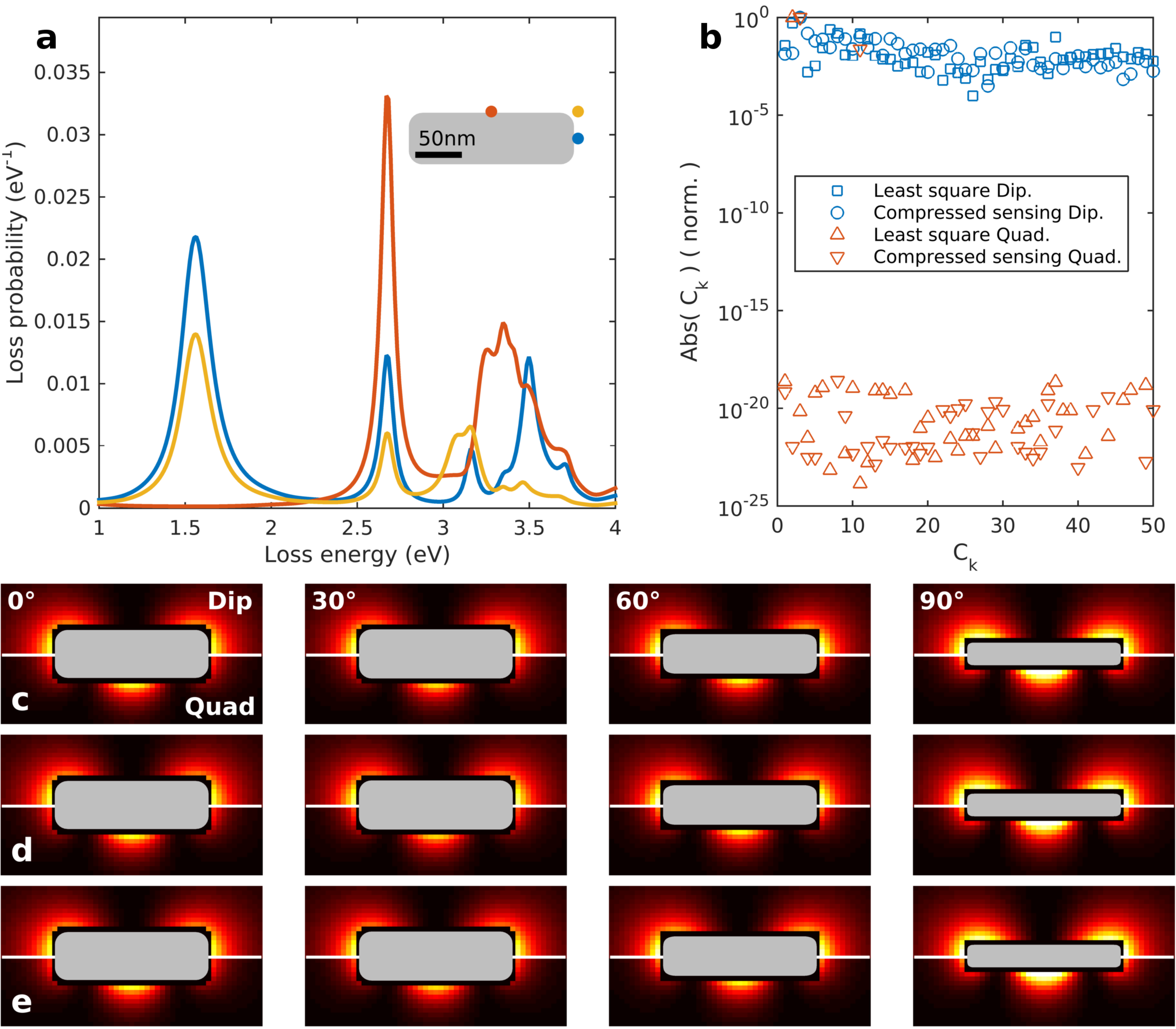}}
\caption{EELS spectra and maps for a silver nanorod.  (a) EELS spectra recorded at the positions indicated in the inset.   The peaks at approximately 1.5 eV and 2.7 eV are attributed to the dipole and quadrupole plasmon mode.  (b) Mode decomposition of the dipole and quadrupole mode from the collection of rotated EELS maps, using either the least square minimization of Eq.~\eqref{eq:optim} or the compressed sensing optimization of Eq.~\eqref{eq:compressed}.  For each mode the coefficients $C_k$ are normalized to unity.  (c) Selected EELS maps for dipole (upper part) and quadrupole (lower part) mode and for different electron propagation directions (rotation angles), as computed with the MNPBEM toolbox.~\cite{hohenester.cpc:12,hohenester.cpc:14b}  (d) Back projected EELS maps for the $C_k$ distribution obtained from the compressed sensing optimization, using Eq.~\eqref{eq:expansion} for the Green function decomposition and Eq.~\eqref{eq:surf2} for the calculation of the loss probabilities.  (e) Same as panel (d) but for $C_k$ distribution obtained from the least square optimization.}
\end{figure}

To prove the applicability of our reconstruction scheme, we generate the ``experimental'' EELS data $\Gamma_{\rm exp}$ using the simulation toolbox MNPBEM for plasmonic nanoparticles.~\cite{hohenester.cpc:12,hohenester.cpc:14b}  We first consider a silver nanorod with dimensions $200\!\times\! 65\!\times\! 30$ nm$^3$ and compute the loss spectra for the three selected impact parameters indicated in Fig.~1a.  The two prominent loss peaks at low energies can be attributed to the dipole and quadrupole plasmon modes.  Corresponding EELS maps at the resonance frequencies are shown for a few selected electron propagation directions (rotation angles) in Fig.~1c.  The mode profiles are reminiscent of the dipole and quadrupole surface charge distributions.~\cite{hoerl.prl:13}  For the decomposition of Eq.~\eqref{eq:expansion} into modes $\bm E_k(\bm r,\omega)$, we use the information about the nanoparticle shape, which in experiment can be obtained from additional high-angle annular dark-field (HAADF) data,~\cite{midgley:09,haberfehlner:14} and compute the $50$ natural oscillation modes of lowest energy (see Methods).  Fig.~1b shows the modulus of coefficents $C_k$ obtained from either a compressed sensing or least square optimization.  Although the two approaches give quite different $C_k$ distributions, the back-projected EELS maps, obtained by assembling the dyadic Green tensor using Eq.~\eqref{eq:expansion} and computing $\tilde\Gamma_{\rm surf}$ from Eq.~\eqref{eq:surf2}, both are in almost perfect agreement with the original $\Gamma_{\rm exp}$ maps.

\begin{figure}
\centerline{\includegraphics[width=\columnwidth]{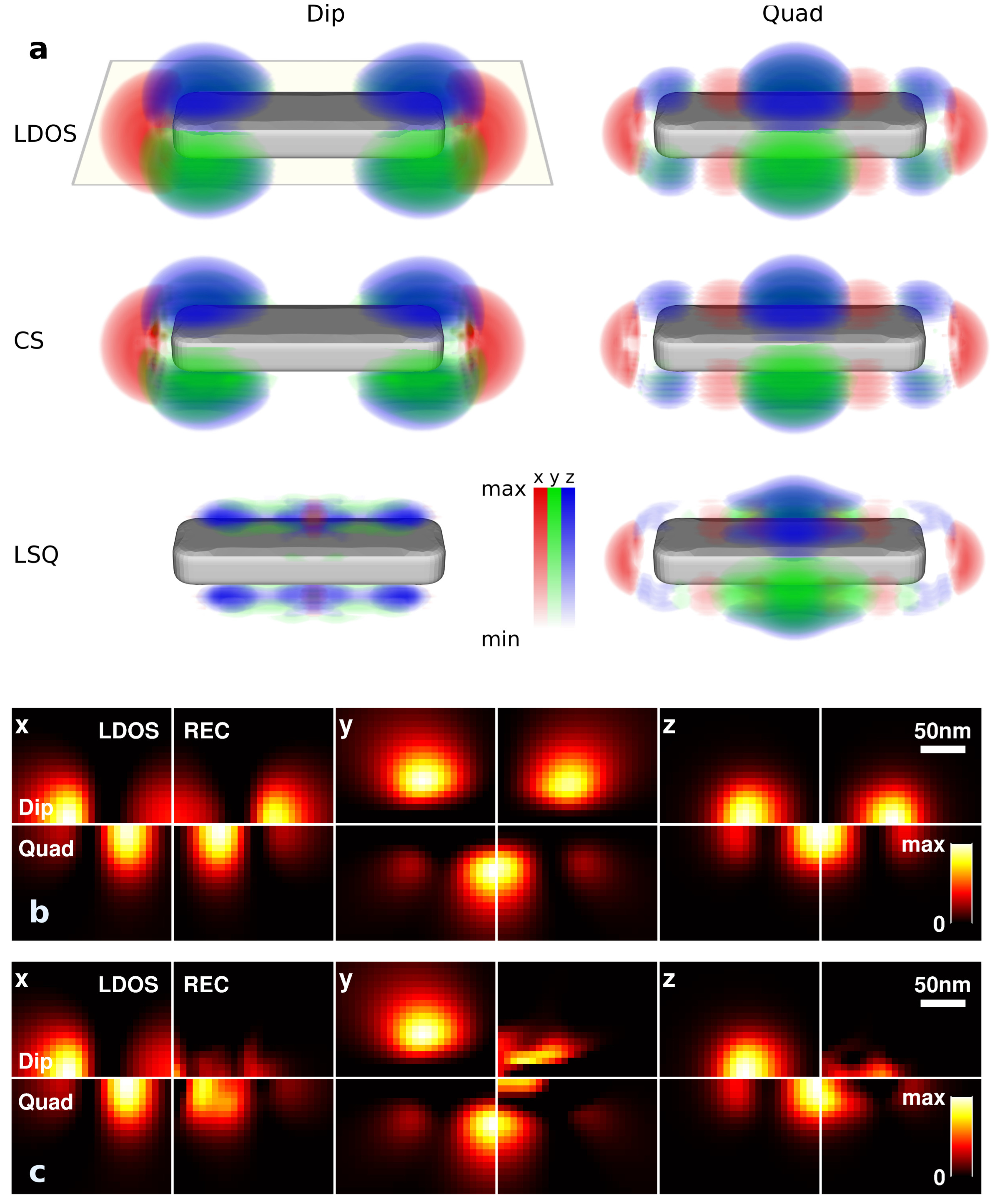}}
\caption{Photonic LDOS of Eq.~\eqref{eq:ldos} and reconstructed LDOS.  (a) Three-dimensional LDOS distribution, as computed with the MNPBEM toolbox (LDOS),~\cite{hohenester.cpc:12} and the distributions reconstructed from the compressed sensing (CS) and least square (LSQ) optimizations.  The projected LDOS $\rho_{\hat{\bm n}}(\bm r,\omega)$ is shown for different projection directions $\hat{\bm n}=\hat{\bm x}$, $\hat{\bm y}$, $\hat{\bm z}$.  (b) LDOS density map in a plane 20 nm above the nanoparticle, as reconstructed from the compressed sensing optimization.  The lower (upper) part of each panel shows the dipole (quadrupole) mode, the left (right) part shows the true (reconstructed) LDOS. (c) Same as panel (b) but for least square optimization.  The reconstructed least square LDOS has also negative contributions, which are set to zero for clarity.}
\end{figure}

Having obtained the $C_k$ values from the optimizations of Eqs.~(\ref{eq:optim},\ref{eq:compressed}) we can use Eq.~\eqref{eq:expansion} to approximately reconstruct the dyadic Green tensor which allows us to compute any electrodynamic response function for the plasmonic nanorod.  In the following we consider the projected photonic LDOS of Eq.~\eqref{eq:ldos}.  Fig.~2 shows the true and reconstructed LDOS maps, and compares the quality of compressed sensing and least square optimizations.  In particular the inspection of panels (b) and (c), which report the LDOS in a plane 20 nm above the nanorod, reveals that the compressed sensing results are in very good agreement with the true LDOS values, whereas the least square optimization completely fails to provide even qualitative agreement.  This finding seems at first sight surprising since both optimization approaches were previously capable of reconstructing the experimental EELS data almost perfectly, as shown Figs.~1c--e.  We attribute the least square shortcoming to the fact that the EELS loss of Eq.~\eqref{eq:surf2} is governed by the long-range tails of the particle plasmon field distributions, with which the passing electron predominantly interacts, whereas the LDOS of Eq.~\eqref{eq:ldos} is governed by the short-range evanescent field components.  Thus, when the optimization has no strong bias on the $C_k$ determination it comes up with the proper long-range components, resulting in high-quality EELS maps shown in Fig.~1e, but fails for the short-range components which contribute little to the minimization function of Eq.~\eqref{eq:optim}.  In contrast, the compressed sensing optimization of Eq.~\eqref{eq:compressed} seeks for a $C_k$ distribution with as few non-zero components as possible.  For suitable basis functions $\bm E_k$ this bias helps to properly select those modes which contribute little but still noticeably to the loss probability of Eq.~\eqref{eq:surf2}.  We emphasize that such a bias for selecting a sparse expansion distribution is by no means unique to the problem of our present concern, but has been previously highlighted in various studies, e.g. in the context of plasmon tomography~\cite{nicoletti:13} or single-pixel cameras,~\cite{duarte:08} and lies at the heart of the compressed sensing algorithm.

\begin{figure}[t]
\centerline{\includegraphics[width=\columnwidth]{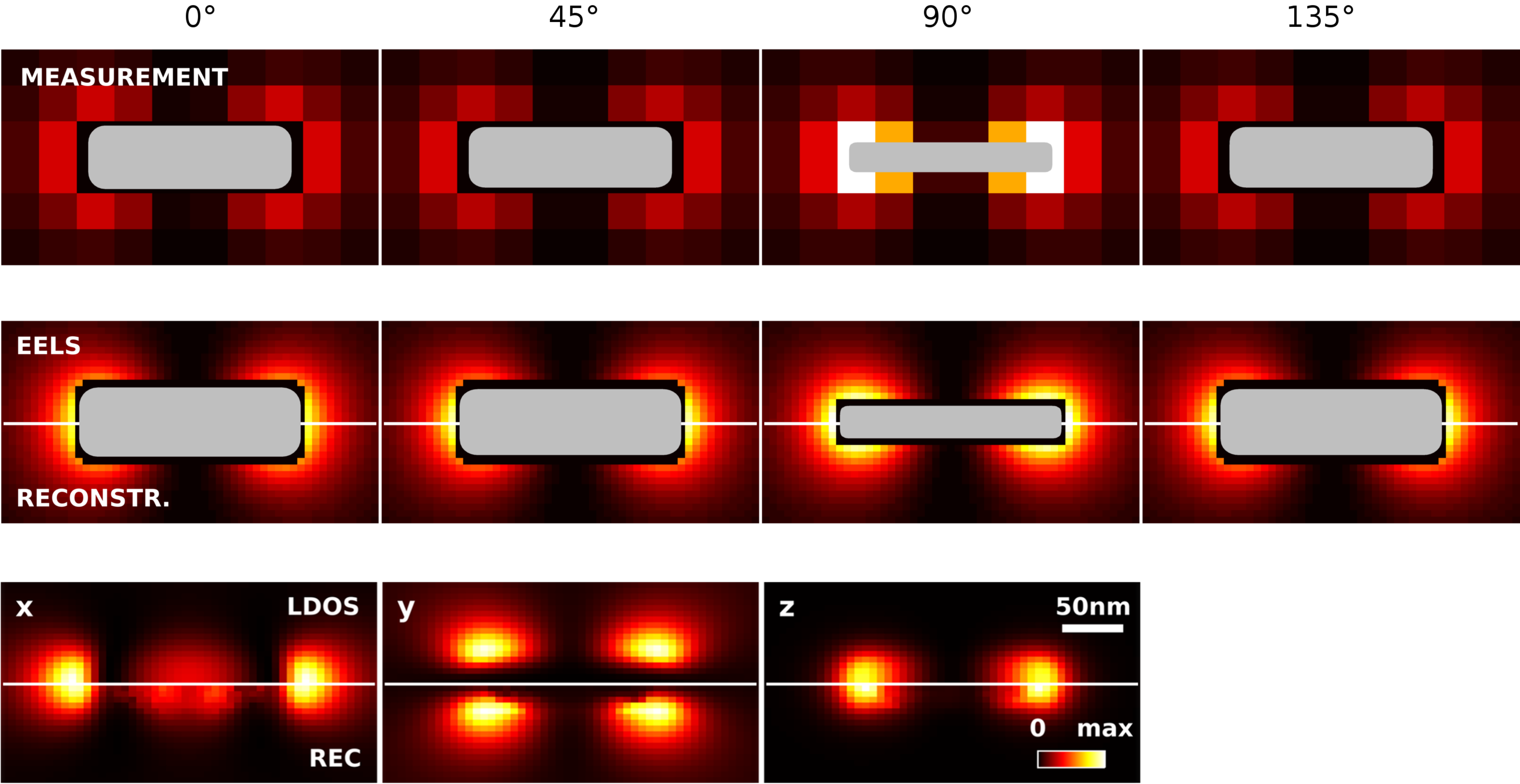}}
\caption{Compressed sensing reconstruction for a strongly reduced number of measurement points.  The first row shows the measurement data for a few rotation angles.  In the second row we compare the EELS data for a finer sampling mesh (upper part of panel) with the reconstructed signal (lower part), finding almost perfect agreement.  The last row reports the true (upper part of panel) and reconstructed (lower part) LDOS maps in a plane 20 nm above the nanorod.}
\end{figure}

An advantage of compressed sensing is that the reconstruction can in general be performed even with a very limited amount of measurement data, and the quality of the reconstructed data is usually not strongly affected by noise.~\cite{candes:08}  In Fig.~3 we show reconstructed EELS and LDOS maps for the small number of impact parameters and rotation angles shown in the first row of measurement data.  As can be seen, the quality of the reconstructed data is extremely good despite the limited amount of measurement data.  This might be beneficial for EELS experiments which typically suffer from a limited amount of rotation angles (missing wedge problem) and where the number of measurement points is often kept low to avoid sample contamination.

\begin{figure}
\centerline{\includegraphics[width=\columnwidth]{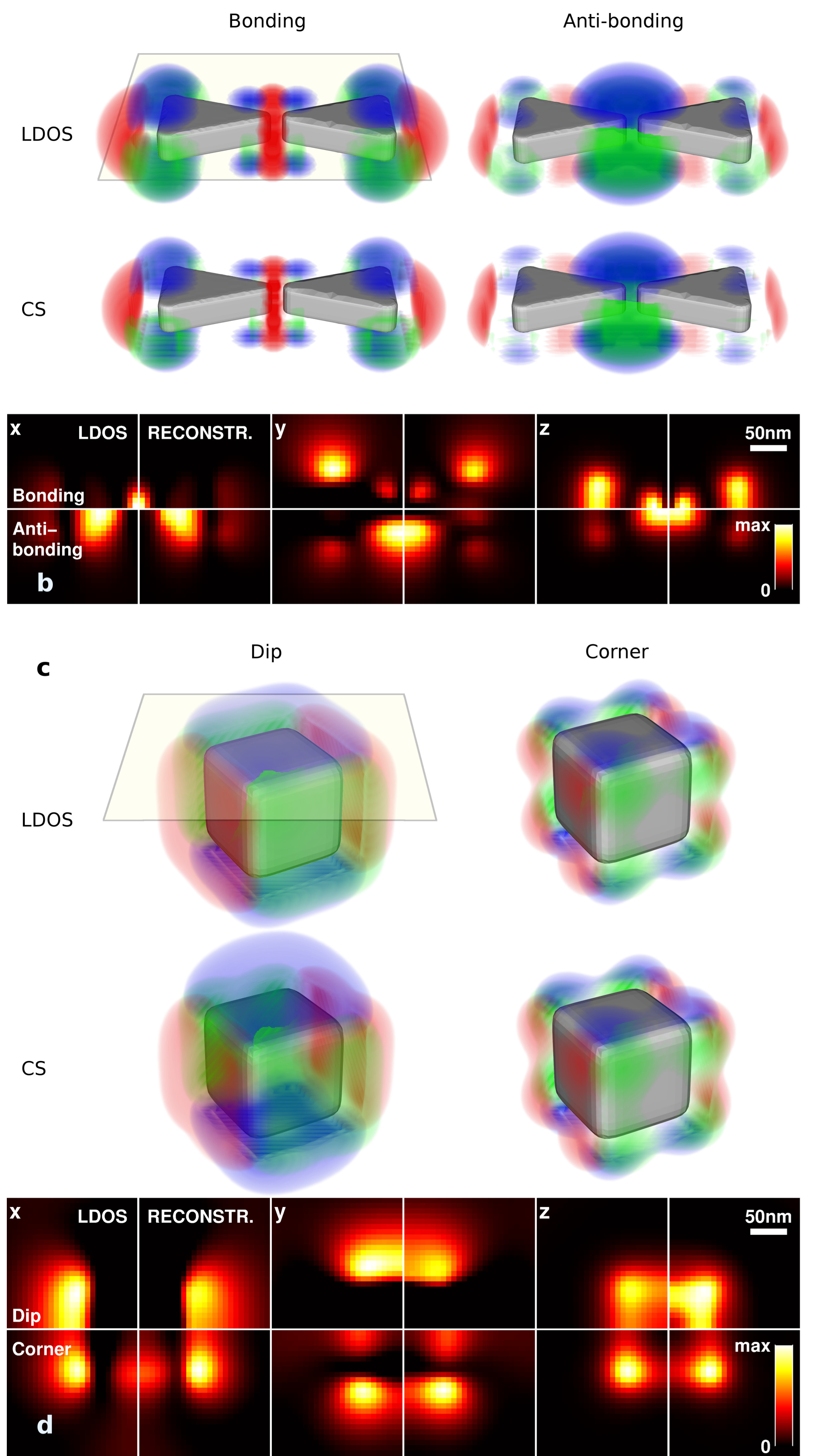}}
\caption{(a) True (upper row) and reconstructed (lower row) LDOS for a bowtie geometry (total size $215\!\times\!85\!\times\!30$ nm$^3$ and 10 nm gap) and for the bonding and anti-bonding modes of lowest energy.  Color code is identical to Fig.~2.  (b) Density map of LDOS in a plane 20 nm above the bowtie structure.  (c) True (left) and reconstructed (right) LDOS for a cube with 150 nm side length, and for the dipole and corner modes of lowest energy.~\cite{nicoletti:13}  (d) Density map of LDOS in a plane 30 nm above the cube.}
\end{figure}

Finally, in Fig.~4 we compare LDOS maps with reconstructed maps for a (a,b) bowtie nanoparticle and (c,d) cube.  For the bowtie geometry we show the LDOS for the two plasmon modes of lowest energy, which can be labelled as bonding and anti-bonding according to the parallel and antiparallel orientation of the dipole moments of the individual nanotriangles.~\cite{schmidt:14b}  The agreement between the true and reconstructed LDOS maps is very good, in particular one can clearly observe the strongly increased LDOS enhancement in the gap region.  For the cube we show the dipole and corner modes of lowest energy,~\cite{nicoletti:13} finding fair agreement between the true and reconstructed LDOS maps.  We attribute the small differences to problems of our algorithm when dealing with degenerate modes of symmetric particles, which might be improved by explicitly accounting for mode symmetries.~\cite{langbein:76}


\section*{Summary and discussion}

To summarize, we have shown how to extract the dyadic Green tensor of Maxwell's theory from a collection of EELS maps recorded for different electron propagation directions (rotation angles).  Our reconstruction scheme is based on a singular-value decomposition of the Green tensor and a compressed-sensing optimization for the expansion coefficients.  We have demonstrated the applicability of our approach for various elementary nanoparticle shapes.  We foresee several improvements for plasmon tomography based on EELS.  On the experimental side, electron holography~\cite{midgley:09} can provide additional information and could allow to disentangle the excitation and measurement channels of plasmonic EELS. On the theoretical side, the presented reconstruction scheme works surprisingly well for most nanoparticle geometries, but further work is needed to clarify the role of various ingredients. 

First, there are several possibilities for chosing the basis functions for the decomposition of the dyadic Green tensor, Eq.~\eqref{eq:expansion}.  In this work we have chosen biorthogonal ``constant flux states''~\cite{tuereci:06} that are the eigenstates of the Green function evaluated for real frequencies (see Methods).  They have the advantage that they can be computed rather straightforwardly, even in case of degenerate or near-degenerate modes, on the other hand they have to be computed for each loss energy separately and several of these modes can govern the plasmonic response.  Another possibility for a basis are the quasi normal modes evaluated at the poles of the Green function in complex frequency space.~\cite{sauvan:13,ge:14,maekitalo:14,alpeggiani:15}  The computation of these modes requires an iterative solution scheme,~\cite{alpeggiani:15} however, once they are computed they can be used for a large frequency range and in general the plasmonic response is only governed by very few of these modes.

In this work we have considered the situation where the basis is already computed for the true nanoparticle shape, and have shown that even in this case the EELS tomography scheme can be quite tricky.  However, our approach is less restrictive than it may appear:  in principle, for electron beams not penetrating the nanoparticle any basis with modes being solutions of the free-space Maxwell's equations can be employed.  Thus, even if a slightly different particle shape or dielectric material is considered in the computation of the basis, this will not necessarily degrade the quality of the reconstruction.  In this case it might be beneficial to adapt our approach such that (i) the modes for the Green function decomposition are expanded in a given non-ideal basis, and (ii) the compressed sensing algorithm seeks for a minimum number of decomposition modes.  Here it might be advantageous to use quasi normal modes, because the same few modes could be optimized for a whole range of loss energies, thus imposing stronger restrictions in comparison to an independent optimization at individual loss energies.

Although further work is needed to establish EELS tomography of plasmonic nanoparticles as a robust and out-of-the-box scheme, we believe that our present work provides an important step forwards for reconstructing electrodynamic quantities from EELS measurements, and makes significant progress with respect to the recently developed tomography schemes that were bound to quasistatic approximation and other restrictive assumptions.

\section*{Methods}

\textbf{Simulations.}  In our simulation approach we compute the LDOS and EELS spectra using the MNPBEM toolbox~\cite{hohenester.cpc:12,hohenester.cpc:14b} and a silver dielectric function extracted from optical experiments.~\cite{johnson:72}  

\textbf{Mode decomposition.}  For the mode decomposition of Eq.~\eqref{eq:expansion} we follow the prescription of Garc\'\i a de Abajo et al.~\cite{garcia:02} and compute the natural oscillation modes through diagonalization of the $\Sigma$ matrix, see Eq.~(21) of Ref.~\citenum{garcia:02} for details, keeping for the solution of the inverse problem the 50 modes of lowest energy.  A higher number of modes didn't show a significant improvement in the reconstruction results.  For our mode decomposition it turns out to be convenient to use a biorthogonal basis, similarily to the quasistatic case.~\cite{boudarham:12}  Our approach closely follows recent related work,~\cite{alpeggiani:15} and we introduce the right and left eigenmodes $\bm E_k(\bm r,\omega)$ and $\tilde{\bm E}_k(\bm r',\omega)$ associated with the $\Sigma$ matrix, respectively.  Instead of the decomposition of Eq.~\eqref{eq:expansion} we then use
\begin{displaymath}
  \bm G(\bm r,\bm r',\omega)\approx\sum_{k=1}^n C_k\, \bm E_k(\bm r,\omega)\otimes\tilde{\bm E}_k^*(\bm r',\omega)\,,
\end{displaymath}
and accordingly also modify  Eq.~\eqref{eq:surfn}.  The biorthogonal expansion turns out to be advantageous in particular for nanoparticles with degenerate modes, as it automatically guarantees proper mode orthonormalization.

\textbf{Compressed sensing.}  The least square optimization is performed with the built-in Matlab functions, for the compressed sensing optimization we use the \texttt{YALL1} software freely available at \texttt{http://yall1.blogs.rice.edu/}.  We set the mixing parameter $\mu=5\times 10^{-2}$ and the stopping tolerance has a value of $10^{-4}$.  We take twelve rotated EEL-maps for each structure with equidistant angles between 0 and 180$^\circ$, each map consisting of $31\!\times\!51$ points. To speed up the optimization process we take only 2000 random measurement points of the generated maps. Further, only measurement points with distance more than 15nm away from the particle surface are used for optimization.  For the volume visualization of the LDOS we use the \texttt{MatVTK} software freely available at \texttt{http://hdl.handle.net/10380/3076}.

\section*{Acknowledgments}

We thank Georg Haberfehlner and Gerald Kothleitner for most helpful discussion.  This work has been supported by the Austria Science Fund (FWF) Projects No. P24511--N26, P27299--N27, SFB F49 NextLite, and NAWI Graz.


\end{document}